\documentclass{emulateapj}

\usepackage{apjfonts}
\usepackage{amsmath}
\usepackage{amssymb}
\usepackage{natbib}
\usepackage{xspace}
\usepackage{graphicx}
\usepackage{rotate} 
\usepackage{calc}
\usepackage{float}

\def\suzaku{\mbox{\instrument{Suzaku}}\xspace}

\newcommand{\err}[2]{\ensuremath{^{+#1}_{-#2}}\xspace}

\newcommand{\Msun}{\ensuremath{M_\odot}\xspace}

\def\kev{ke\kern -0.05em V\xspace}

\newcommand{\instrument}[1]{\textsl{#1}\xspace}

\slugcomment{To be submitted.}

\shorttitle{1A\,1118$-$61 with Suzaku}
\shortauthors{Suchy et al.}

\begin{document}
\title{Suzaku observations of the HMXB 1A\,1118$-$61} 

\author{Slawomir~Suchy\altaffilmark{1},
Katja~Pottschmidt\altaffilmark{2,3},
Richard~E.~Rothschild\altaffilmark{1},
J\"orn~Wilms\altaffilmark{4.5},
Felix~F\"urst\altaffilmark{4,5},
Laura~Barragan\altaffilmark{4,5},
Isabel~Caballero\altaffilmark{6},
Victoria~Grinberg\altaffilmark{4,5},
Ingo~Kreykenbohm\altaffilmark{4,5},
Victor~Doroshenko\altaffilmark{7},
Andrea~Santangelo\altaffilmark{7},
R\"udiger~Staubert\altaffilmark{7},
Yukikatsu~Terada\altaffilmark{8},
Wataru~Iwakari\altaffilmark{8},
Kazuo~Makishima\altaffilmark{9,10} 
}

\altaffiltext{1}{University of California, San Diego, Center for
  Astrophysics and Space Sciences, 9500 Gilman Dr., La Jolla, CA
  92093-0424, USA}
\altaffiltext{2}{Center for Space Science and Technology, University of Maryland Baltimore County, 1000 Hilltop Circle, Baltimore, MD 21250, USA}
\altaffiltext{3}{CRESST and NASA Goddard Space Flight Center, Astrophysics Science Division, Code 661, Greenbelt, D 20771, USA}
\altaffiltext{4}{Dr. Karl Remeis Sternwarte, Astronomisches Institut,
  Sternwartstr. 7, 96049 Bamberg, Germany} 
  \altaffiltext{5}{Erlangen Centre for Astroparticle Physics, University of Erlangen-Nuremberg, Erwin-Rommel-Strasse 1, 91058 Erlangen, Germany}
\altaffiltext{6}{CEA Saclay, DSM/IRFU/SAp -UMR AIM (7158) , CNRS/CEA/Univ. P.Diderot -F-91191 Gif sur Yvette France}
\altaffiltext{7}{Institut f\"ur Astronomie und Astrophysik 
  Astronomie, Sand 1, 72076 T\"ubingen, Germany} 
\altaffiltext{8}{Graduate School of Science and Engineering, Saitama University,
255 Simo-Ohkubo, Sakura-ku, Saitama city, Saitama 338-8570, Japan}
\altaffiltext{9} {Department of Physics, Graduate School of Science,
University of Tokyo, Hongo 7-3-1, Bunkyo-ku, Tokyo 113-0033, Japan}  
\altaffiltext{10}{High Energy Astrophysics Laboratory, 
Institute of Physical and Chemical Research (RIKEN), 
Hirosawa 2-1, Wako, Saitama 351-0198, Japan}   
  
\email{ssuchy@ucsd.edu}


\begin{abstract} 
We present broad band analysis of the Be/X-ray
transient 1A\,1118$-$61 by \textsl{Suzaku} at the peak of its 3rd observed outburst in January 2009
and 2 weeks later when the source flux had decayed by an order of magnitude. The continuum was
modeled with a \texttt{cutoffpl} model  as well as a \texttt{compTT} model, with both cases requiring an additional
black body component at lower energies.  We confirm the detection of a
cyclotron line at $\sim$~55\,keV and discuss the possibility of a
first harmonic at $\sim$~110\,keV. 
Pulse profile comparisons show a change in the profile structure at lower energies, 
an indication for possible changes in the accretion geometry.
Phase resolved spectroscopy in the outburst data show a change in the continuum
throughout the pulse period. The decrease in the CRSF centroid energy also indicates that the viewing angle 
on the accretion column is changing throughout the pulse period. 
\end{abstract}

\section{Introduction}\label{sec:introduction}
The Be/X-ray binary transient 1A\,1118$-$61 was serendipitously
discovered during an observation of the nearby binary system Cen\,X$-$3,
when an outburst was detected in December of 1974 by the
\textsl{Ariel-5} satellite \citep{Eyles:1975}.  A second, similar
outburst occured in January of 1992 and was observed by the Burst and 
Transient Source Experiment on the Compton Gamma Ray observatory 
\textsl{CGRO}/BATSE \citep{Coe:1994}. The measured peak flux was $\sim
150$ mCrab for the 20-100\,keV energy range, similar to the 1974
outburst.  The source showed an elevated emission throughout the next
$\sim 30$\,days after the decay of the main outburst
\citep[see][Fig.\,1] {Coe:1994}. The third and most recent outburst
occurred on 2009, January 4 and was detected by the \textsl{Swift} Burst Alert Telescope BAT
\citep{Mangano:2009, Mangano:2009a}. It reached a peak flux of $\sim
500$\,mCrab in the 15-50\,keV energy band. This last outburst was monitored with \textsl{Swift} and the Rossi X-ray Timing Explorer 
(\textsl{RXTE}) as well as with two long \textsl{Suzaku} pointings
and one observation with \textsl{INTEGRAL} during a flaring episode $\sim 50$\, days
after the peak of the main outburst \citep{Leyder:2009}.

Pulsations with a period of $ 405.3 \pm 0.6$\,s were observed during
the 1974 outburst and were initially attributed to the orbital period
of two compact objects \citep{Ives:1975}. Shortly afterwards it was
suggested that the period stems from a slow rotation of the neutron star (NS)
itself \citep{Fabian:1975}. During the 1992 outburst pulsations with
a period of $\sim 406.5$\,s were detected up to 100\,keV, showing a
broad, asymmetric, single peak pulse profile above the lowest BATSE
energy of 20\,keV. The pulse period decreased throughout the
decline of the outburst with a rate of $\sim -0.016$\,s/day and it
appeared constant at $\sim 406.35$\,s for the time of the
elevated emission. During the 2009 outburst a similar period evolution was
observed with \textsl{RXTE}
resulting in a pulse period of $P_\text{spin} = 407.719$\,s, and
$\dot{P}_\text{spin} = -4.6\times 10^{-7} $\,s/s $\approx -0.04$\,s/day
\citep{Doroshenko:2010}. Furthermore, the lower energies showed a more
complex pulse profile with two peaks below an energy of $\sim10$\,keV. Due
to the short duration of the \textsl{Suzaku} observations with respect
to the pulse period, the derived \textsl{RXTE} values are used for determining the
pulse profile and phase resolved spectra in this paper.

The optical counterpart was identified as the Be-star Hen
3$-$640/Wray~793 by \citet{Chevalier:1975} and classified as an
O9.5IV-Ve star with strong Balmer emission lines indicating an
extended envelope by \citet{Janot-Pacheco:1981}.  The overall spectrum
was found to be similar to other known Be/X-ray transients, such as
X$-$Per and A\,0535+26 \citep[][and references therein]{Villada:1992}. The
distance was estimated to be $5\pm2$\,kpc \citep{Janot-Pacheco:1981}
and was confirmed by
\citet{Coe:1985}, along with the spectral type classification, using UV observations of the source.
 \textsl{EXOSAT} observed X-ray emission from
1A\,1118$-$61 between outbursts \citep{Motch:1988}, thus indicating a continuous low
level of accretion. \citet{Rutledge:2007}
reported on pulsations in the low luminosity state observed with
\textsl{Chandra}, making it only the third known HMXB transient after
A\,0535+26 and 4U\,1145-619 for which this behavior has been observed. 

A study of the $H_\alpha$ emission line before and during the 1994 outburst \citep{Coe:1994} showed
a strong correlation between its strength and the observed X-ray
activity, indicating the existence of a very large disk around the
star. The fluctuation of the equivalent width of $H_\alpha$ indicated 
possible instabilities in the disk, which were enhanced when the star passes
through the periastrion. The analysis of the UV continuum and line
spectra has indicated that the photospheric emission from the Be star was
not affected by the X-ray radiation, similar to the case of A\,0535+26
\citep{deloore:1984}.

Until recently, the orbital period of 1A\,1118$-$61 was not measured and assumed values
were of the order of 350\,days, based on the $P_{spin} - P_{orb}$ relation \citep{Corbet:1986}, and 
$\sim 585$\,days based on the equivalent width of the $H_\alpha$ line \citep{Reig:1997}. 
\citet{Staubert:2010} analyzed the pulse arrival time in \textsl{RXTE} data 
throughout the 2009 outburst and established an orbital period of $\sim 24$\,days with a 
very circular orbit around the Be-companion. 

Spectral fitting during the 1974 outburst indicated a variable 
power law index, where the lowest value was $\sim 0.9$ during the peak
of the outburst, and $\sim 1.1$ before and after
\citep{Ives:1975}. \citet{Coe:1994} described the 1992 \textsl{CGRO}/BATSE data
with a single-temperature, optically-thin, thermal bremsstrahlung
(OTTB) with a temperature of 15\,keV. Due to the different energy
ranges of the instruments and the fact that pulse$-$off-pulse data have been used,
these results can not be directly compared with the 1974 observation.

In the 2009 outburst \citet{Doroshenko:2010}
discovered a cyclotron resonance scattering feature (CRSF) at $\sim
55$\,keV.  CRSFs can be used to deduce the strength of the magnetic field
at the pole region of the NS and have been observed in
multiple sources at energies from $\sim 15$\,keV up to $\sim 50$\,keV
\citep{Coburn:2002}. 

\textsl{Suzaku} observed 1A\,1118$-$61 twice in 2009, once during the
peak of the outburst and again $\sim 13$\,days later when the flux had
returned to a level slightly above the quiescent state. In this paper
we will present the first detailed 0.7-200\,keV broad band spectrum
for this source, confirming the detection of a CRSF at $\sim
55$\,keV. The paper is divided as follows:
Section 2 describes the data accumulation and extraction; Section 3
presents the broad band phase averaged spectra for the outburst
(0.7--200\,keV) and for the second observation (0.7$-$70\,keV), and
discusses the CRSF and
Fe-Line complex; Section 4 compares energy resolved pulse profiles for both
observations and discusses the phase resolved analysis for the
outburst data; finally Sections 5 and 6 present a discussion of our observations and
summarize our findings.

 \begin{figure}
 \plotone{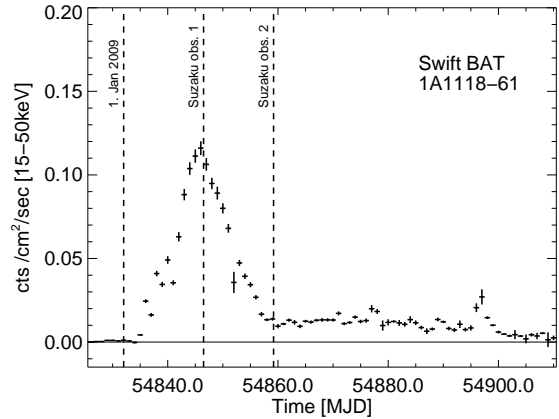}
\caption{\textsl{Swift}/BAT lightcurve during the 1A\,1118$-$61 outburst. The dashed lines mark the Suzaku observation as well as 2009 Jan 1.}
\label{fig:light}
\end{figure}

\section{Observation and Data Reduction}\label{data}
A sudden increase in activity of 1A\,1118$-$61 was detected with the
\textsl{Swift}/BAT instrument on January 4th, 2009 \citep{Mangano:2009}.  
The count rate
increased steadily until January 15th when it reached the maximum
value of $\sim 500$\,mCrab and then decayed until January 27th where it
showed a low emission level with periods of flaring \citep{Leyder:2009} until mid March and then returned to
quiescence (see Fig.~\ref{fig:light}).  \suzaku observed 1A\,1118$-$61 during the peak of the
outburst on January 15th, 2009 (MJD 54846.5, ObsID 403049010) with
both of its main instruments: the X-ray Imaging Spectrometer
\citep[XIS;][]{Mitsuda:2007} and the Hard X-ray Detector
\citep[HXD;][]{Takahashi:2007}. A second observation was performed on
January 28th, 2009 (MJD 54859.2, ObsID 403050010), $\sim 13$ days after the main
outburst, at the beginning of the period of elevated emission. 

Both observations were
performed using the HXD nominal pointing to minimize the pile-up fraction in
the XIS instruments and to enhance the HXD sensitivity for a possible
CRSF detection.  Of the four original XIS instruments only XIS 0, 1 and
3 were functional during the observing time.  XIS 0 and 3 are both
front illuminated (FI) CCDs and XIS 1 is a back illuminated (BI) CCD.
To reduce pile-up, the XIS instruments were operated with the 1/4
window option with a readout time of 2\,s and the burst option with
only 1\,s accumulation time for each readout cycle, reducing the exposure
time for the XIS instruments by 50\%.  The data were taken in the $3
\times 3$ and $5\times 5$ editing modes which were then combined for the 
final spectral analysis. 

For the extraction the \textsl{Suzaku} FTOOLS
version 16 (part of HEASOFT 6.9) was used. The unfiltered XIS
data were reprocessed with caldb20090402 and then screened with the standard
selection criteria as described in the ABC
guide\footnote{http://heasarc.gsfc.nasa.gov/docs/suzaku/analysis/abc/}. Each
detector and editing mode combination was extracted independently and
individual response matrices and effective area files were
created. For the final spectra the data from both FI detectors were
combined (XIS 0 and 3) and the response matrices and effective areas were
weighted according to the accumulated exposure time of the different
modes. The XIS data were grouped so that the minimum number of
channels per energy bin corresponded to at least the half width half
maximum of the spectral resolution, i.e. grouped by 8, 12, 14, 16, 18, 20, 22
channels starting at 0.5, 1, 2, 3, 4, 5, 6, and 7\,keV, respectively
(Nowak, priv. com).

For the HXD data the \textsl{Suzaku} team provided the tuned PIN non
X-ray
background\footnote{http://heasarc.nasa.gov/docs/suzaku/analysis/pinbgd.html}
(NXB).  Following the ABC Guide the cosmic X-ray background (CXB) was
simulated and  the exposure time was adjusted in both backgrounds
by the prescribed factor of 10. The PIN data were grouped so that at least 100
events were detected in each spectral bin.

GSO data were extracted using the FTOOL \texttt{hxdpi} with the newest
gain calibration file from April, 2010. The NXB background files
version 2.4 created by the \textsl{Suzaku} HXD instrument team were
used. The data were then binned to 64 bins according to the Suzaku
homepage\footnote{http://heasarc.gsfc.nasa.gov/docs/suzaku/analysis/gso\_newgain.html}. GSO
data in the range 70$-$200\,keV were used in the spectral analysis.

These selection criteria resulted in $\sim$ 25\,ks exposure time for each XIS instrument and $\sim$ 49\,ks for the HXD instruments in the first observation. 
The second observation had an exposure of $\sim$ 21\,ks and $\sim$ 29\,ks for the XIS and HXD instruments, respectively. 

\subsection{Pileup correction}
For bright sources, such as 1A\,1118$-$61 during the outburst, a strong
pileup is expected. Following the description provided at
\texttt{http://space.mit.edu/ASC/software/suzaku/} the S-lang routine
\texttt{aeattcor.sl} was used to improve the attitude correction file
and the point spread function of the events. Then the tool
\texttt{pile\_estimate.sl} was applied to produce a 2 dimensional map
of the pileup fraction. The maximum values for pileup fractions were 10\% and
15\% for XIS0, 15\% and 16\% for XIS1, and 18\% and 21\% for XIS3 for
the $3\times3$ and $5\times5$ editing modes, respectively. Regions
with a pileup fraction above 5\% were excluded from the extraction for
each individual source event file.
For the second observation the calculated maximum pileup fractions were < 5\% 
and no regions had to be excluded for the extraction. 

\section{Phase averaged spectrum} 
For the outburst observation we extracted broad band XIS spectra for
0.7$-$12\,keV (0.7$-$10\,keV for the BI XIS1), PIN spectra for 12--70\,keV
and GSO spectra for 70--200\,keV.  For the second observation the GSO
spectrum was not well constrained and therefore was not included in the
analysis. The final model included the Galactic and intrinsic absorption, a continuum 
that steepened at higher energies, an iron line complex, and a CRSF. 
Excessive low energy flux was modeled with a black body component. In addition a 10\,keV 
absorption feature was required in the outburst observation. 

The Galactic column density was modeled with a single photon
absorption (\texttt{phabs}) component, where the column density was
confined between 1.1 and 1.4 $\times 10^{22}$\,cm$^{-2}$. 
The value determined by the NASA $N_\text{H}$
Tool\footnote{http://heasarc.gsfc.nasa.gov/cgi-bin/Tools/w3nh/w3nh.pl}
for 1A\,1118$-$61 is $1.22\times10^{22}$\,cm$^{-2}$.
The $N_\text{H}$ value was left free in the fits for the first observation. For the second 
observation the values showed larger error bars with the \texttt{cutoffpl} model and had to be fixed when using a \texttt{compTT} model. 
The intrinsic column
density was modeled with the partially covered photon absorption model
\texttt{pcfabs} to take the flux at lower energies (<1\,keV) into
account. All modeling was performed using the \texttt{wilm} abundances
\citep{Wilms:2000} with the \texttt{vern} cross-sections
\citep{Verner:1996}. The continuum was
modeled using a power law with an exponential cutoff
(\texttt{cutoffpl}).  Using a power law with a Fermi-Dirac (FD)
cutoff, one of the empirical models often applied to accretion powered
pulsars, resulted in a cutoff energy of $\sim 0$, making the FD cutoff
effectively a \texttt{cutoffpl}, where the variable $E_\text{cut}$ actually reflects 
the folding energy of the model. The additional 10\,keV feature was modeled using a
negative, broad ($\sigma = 1-2$\,keV) Gaussian component. Such a feature has
been previously observed in different sources and it is believed to
stem from an improper modeling of the continuum \citep[see \S6.4
  in][]{Coburn:2002}. More recent examples of such a feature are,
e.g. 4U1907+09 \citep{Rivers:2010} and Cen X$-$3
\citep{Suchy:2008}. Although this feature sometimes is described as a broad
emission line, e.g. \citet{Suchy:2008}, in this case a  negative Gaussian line at $\sim$10\,keV 
fits best.
When including this line in the overall best fit, 
the $\chi^2$/dofs (degrees of freedom) decreased from 1173 / 477 to the best fit value of 678 / 474, respectively.  
Due to the small differences in the
instrument response, all detectors were coupled with a normalization
constant, which was set to 1 for the combined XIS 03 detectors.
The final model is of the form \texttt{const*phabs*pcfabs*(cutoffpl*gabs+ 2*Gauss$_\text{Fe}$)+Gauss$_\text{10\,keV}$+3*Gauss$_\text{Cal}$}.
The best fit parameters are mentioned in the text and partially summarized in Table~\ref{tab1}.

 \begin{figure}
 \plotone{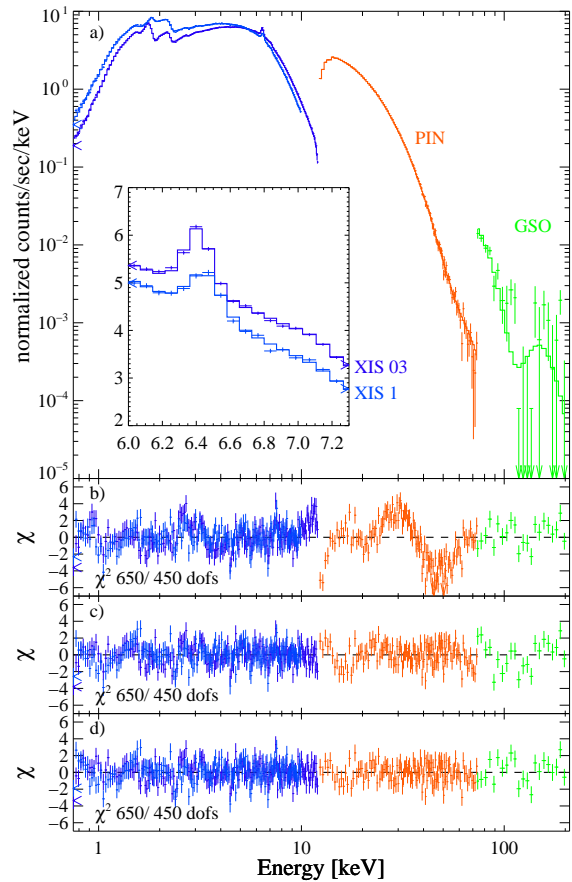}
\caption{Broad band phase averaged spectrum of the outburst
  observation using the \texttt{cutoffpl} model. The residuals are shown without CRSF (b), with one CRSF
  at 55\,keV (c) and with the second CRSF at 112\,keV. The Inset shows
  the Fe line region and the best fit model for XIS1 and XIS 03. }
\label{fig:on_spec}
\end{figure}

 \begin{figure}
 \plotone{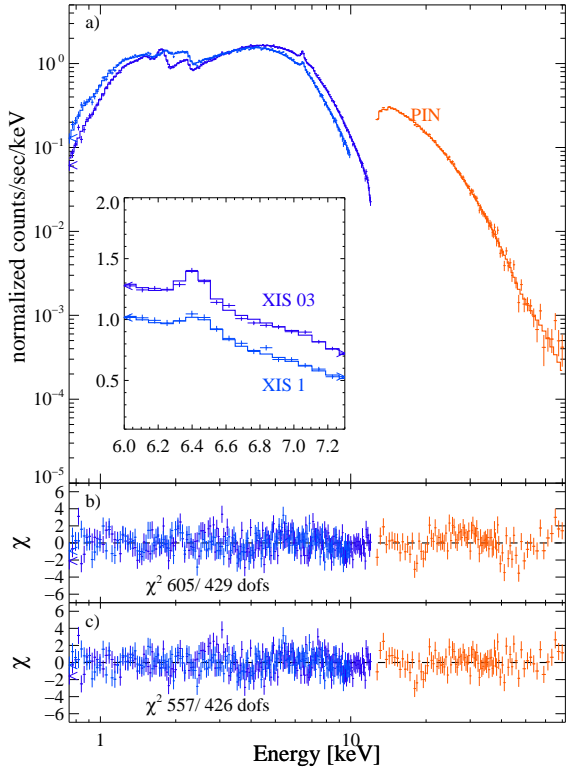}
\caption{Broad band phase averaged spectrum of the second
  observation using the \texttt{cutoffpl} model. Only XIS and PIN data were used. Residuals without a
  CRSF (b) and with one CRSF at $\sim 47$\,keV. The Inset shows the Fe
  line region and the best fit model for XIS1 and XIS 03. }
\label{fig:off_spec}
\end{figure}

\begin{deluxetable*}{lccclcc}
	\tabletypesize{\scriptsize}
	\tablecaption{Phase averaged spectral parameter. Model described in the text. \label{tab1} }
	\tablecolumns{8}
\startdata
\hline
\hline
Parameter  &  \multicolumn{3}{c}{Cutoffpl}  & Parameter &  \multicolumn{2}{c}{CompTT}  \\                 
                                                                                   &  Outburst$^{a}$                      	& 2nd Obs.      &  2nd Obs, free CRSF &&  Outburst                      & 2nd Obs.   \\
\text{phabs} $N_\text{H} [10^{22}$/cm$^2$]      & $1.34^{+0.01}_{-0.01}$ 	& $1.34 $ frozen &  $1.43^{+0.14}_{-0.13}$      		&& $1.28^{+0.03}_{-0.02}$ & $1.28$ frozen\\ 
\text{pcfabs} $N_\text{H} [10^{22}$/cm$^2$] 	& $11.2^{+0.2}_{-0.3}$ 	& $13.5^{+0.4}_{-0.4}$ & $13.6^{+0.5}_{-0.5}$    			&& $12.00^{+0.4}_{-0.5}$ & $15.1^{+0.1}_{-0.7}$\\ 
covering fract.	  						& $0.66^{+0.01}_{-0.01}$ & $0.71^{+0.01}_{-0.01}$ & $0.71^{+0.01}_{-0.01}$       		&& $0.71^{+0.01}_{-0.01}$ & $0.76^{+0.01}_{-0.01}$\\ 
blackbody kT [keV]  						& $0.51^{+0.01}_{-0.003}$& $0.15^{+0.01}_{-0.01}$ &  $0.15^{+0.01}_{-0.01}$      		&& $0.53^{+0.01}_{-0.01}$ & $0.17^{+0.01}_{+0.01}$\\ 
blackbody norm   $[10^{-3}]$ 			& $7.3^{+0.4}_{-0.2}$       	& $2.2^{+0.2}_{-0.03}$ &  $2.85^{+1.67}_{-1.22} $     		&& $12.8^{+1.1}_{-0.9}$ & $2.89^{+0.03}_{-0.02}$\\ 
$E_\text{cut}$ [keV]   					& $18.7^{+0.2}_{-0.2}$& $26.3^{+3.0}_{-2.5}$ & $24.7^{+1.8}_{-1.6}$         &compTT T$_0$ [keV]  & $1.34^{+0.04}_{-0.05}$ & $0.44^{+0.06}_{-0.04}$\\ 
$\Gamma$   							& $0.350^{+0.022}_{-0.004}$ & $1.02^{+0.04}_{-0.04}$ & $1.01^{+0.03}_{-0.03}$ &compTT kT [keV] & $7.6^{+0.3}_{-0.3}$ & $182^{+158}_{-96}$\\ 
$A_\text{pl} [10^{-2}]^\dagger$				& $9.55^{+0.02}_{-0.03}$ & $5.8^{+0.3}_{-0.3}$ & $5.7^{+0.3}_{-0.3}$            &compTT $\tau_\text{p}$ & $6.00^{+0.19}_{-0.18}$ & $0.68^{+1.33}_{-0.26}$\\
									& 					& 						&&	compTT norm$^\dagger$ 	& $0.096^{+0.003}_{-0.003}$ & $0.002^{+0.004}_{-0.001}$\\ 
$E_\text{CRSF}$ [keV] 					& $58.2^{+0.8}_{-0.4}$ & $58.2$ frozen  & $47.4^{+3.2}_{-2.3}$      			&& $54.5^{+2.4}_{-2.1}$ & $54.5$ frozen\\ 
$\sigma_\text{CRSF}$ [keV]				& $14.1^{+3.5}_{-3.1}$ & $14.1$ frozen &  $5.7^{+2.0}_{-1.7}$     			 && $10.3^{+3.6}_{-0.1}$ & $10.3$ frozen\\ 
$\tau_\text{CRSF}$ 						& $60.1^{+5.9}_{-1.5}$ & $15.4^{+6.4}_{-6.3}$ &  $6.0^{+3.0}_{-2.1} $     	&& $23.6^{+10.4}_{-6.9}$ & $27.1^{+3.2}_{-3.0}$\\ 
$E_{\text{Fe~K}_\alpha}$ [keV] 				& $6.414^{+0.006}_{-0.002}$ & $6.42^{+0.01}_{-0.01}$ & $6.42^{+0.01}_{-0.01}$&& $6.42^{+0.01}_{-0.01}$ & $6.42^{+0.01}_{-0.01}$\\ 
$E_{\text{Fe~K}_\beta}$ [keV]      			& $7.13^{+0.03}_{-0.03}$ & $7.13$ frozen  	&7.13 frozen		&& $7.13^{+0.03}_{-0.03}$ & $7.13$ frozen \\ 
Eq. Width K$_\alpha$ / K$_\beta$ [eV] 		&  $51\err{4}{4} / 8\err{3}{3}$  & $46\err{10}{10} / 8\err{2}{2}$                  &$46\err{10}{10}/ 8\err{2}{2}$   &&  $51\err{4}{4} / 10\err{3}{3}$ & $49\err{10}{10}/ 11\err{9}{9} $      \\ 

Flux$_{2-10\,\text{keV}}   [10^{36}$ ergs/sec ]$^\ddagger$  &   8.79 & 1.72    & 1.72 && 9.15 & 1.72 \\
C$_\text{XIS1}$/C$_\text{PIN}$/C$_\text{GSO}$ &  0.98 / 1.10 / 0.82       &  1.02 / 1.01/--&1.02/1.01/--     		&& 0.99/ 1.18 / 1.10 & 0.95/ 1.00/--\\ 
$\chi^2$/dofs     				                     & 678 / 474 		& 574 / 431 &  567 / 428      				&& 680 / 474           & 567 / 428\\ 

\hline
\enddata
\tablecomments{$^{a}$ Best fit values including first harmonic at  $\sim 110$\,keV, see text for details. $\dagger$ Units in Photons keV$^{-1}$ cm$^{-2}$ s$^{-1}$, $\ddagger$ unabsorbed flux using a distance of 5\,kpc}
\end{deluxetable*}

A second approach for the continuum was initiated by
\cite{Doroshenko:2010} using the Comptonization model \texttt{compTT}
developed by \cite{Titarchuk:1995} instead of the \texttt{cutoffpl} model.  The parameters of the
\texttt{compTT} were again left independent between the FI XIS03 and
the BI XIS1, but only small differences could be observed. 
The negative Gaussian at 10\,keV was not necessary for
this model. When comparing the spectral parameters in the outburst between these two models,
i.e. CRSF, $N_\text{H}$ and the spectral lines, to the previously
described \texttt{cutoffpl}, the values were consistent.

A comparison of the \texttt{compTT}
values, such as the photon seed temperature $T_0$, the plasma
Temperature $kT_\text{e}$ or the optical depth $\tau_p$ showed consistency between 
the spectra of the outburst as seen by \textsl{Suzaku} and \textsl{RXTE} \citep{Doroshenko:2010}. 

When comparing the outburst spectrum with the second observation,
significant changes in the continuum parameters are observed. In the
\texttt{cutoffpl} model the power law index softened from
$0.350^{+0.022}_{-0.004}$ to $1.01^{+0.03}_{-0.03}$. The cutoff energy
$E_\text{cut}$ increased from $18.7^{+0.2}_{-0.2}$\,keV during the
outburst to $24.7^{+1.8}_{-1.6}$\,keV in the second observation. The
temperature of the additional black body decreased from
$0.51^{+0.01}_{-0.003} $\,keV to $0.15^{+0.01}_{-0.01}$\,keV after
the outburst. The intrinsic column density and covering fraction increased both from 
$11.2^{+0.2}_{-0.3}\times10^{22}$\,cm$^{-2}$ and $0.66^{+0.01}_{-0.01}$ in the outburst 
to $13.6^{+0.5}_{-0.5}\times10^{22}$\,cm$^{-2}$ and $0.71^{+0.01}_{-0.01}$ in the second observation.

In the \texttt{compTT} model the changes between the two observations were 
similar. The intrinsic column density and black body values were consistent with the 
values obtained with the \texttt{cutoffpl} model, where the intrinsic column density 
is $12.0^{+0.4}_{-0.5}\times 10^{22}$\,cm$^{-2}$ and $15.1^{+0.1}_{-0.7}\times 10^{22}$\,cm$^{-2}$ and the covering fraction is 
$0.71^{+0.01}_{-0.01}$ and $0.76^{+0.01}_{-0.01}$ for the outburst and the second observation, respectively. When establishing the 
errors for the other spectral parameters of the continuum, these values had to be
frozen to avoid a drift into an unreasonable part of the parameter space. 
The black body temperature declined from $0.53^{+0.01}_{-0.01}$\,keV to $0.17^{+0.01}_{-0.01}$\,keV, 
similar values as in the \texttt{cutoffpl} model. 
The \texttt{compTT} values changed significantly between the two observations. 
The photon seed temperature decreased from $1.34^{+0.04}_{-0.05}$\,keV to $0.44^{+0.06}_{-0.04}$\,keV. 
The electron plasma temperature $kT_\text{e}$
increased from $7.6^{+0.3}_{-0.3}$\,keV to $182^{+158}_{-96}$\,keV, whereas the optical depth of the 
plasma $\tau_p$ decreased from $6.00^{+0.19}_{-0.18}$ to $0.68^{+1.33}_{-0.26}$. This behavior reflects the known 
negative $kT_\text{e}$ and $\tau_p$ correlation (see Section 5.2).
 
\subsection{Fe line component} 
An Fe~K$_\alpha$ emission component was detected at
$6.414^{+0.006}_{-0.002}$\,keV in the outburst data (see
Fig. \ref{fig:on_spec}, Inset), as well as at
$6.42^{+0.01}_{-0.01}$\,keV  in the 2nd observation (see Fig. \ref{fig:off_spec}, Inset). For the best fit values 
the centroid energies of XIS03 and XIS 1 were left independent, due to a known 
shift in the energy calibration. 
Although the centroid energies for XIS1 generally show a slightly lower (30\,eV) energy 
than the other XIS instruments\footnote{http://www.astro.isas.jaxa.jp/suzaku/process/caveats/caveats\_xrtxis06.html},  
our XIS1data instrument actually showed a 20-30\,eV higher centroid energy. 
Residuals at $\sim 7$\,keV were observed and were
modeled with an additional Gaussian line with a best-fit energy of $7.13\pm0.03$\,keV,
indicating the existence of an Fe~K$_\beta$ line. 
The width $\sigma$ of the Fe~K$_\beta$ line was set to the  
Fe~K$_\alpha$ line value, and the normalization was left independent. 
The obtained value for the K$_\beta$ normalization was consistent with the expected 12\% of the K$_\alpha$ normalization. 
In the second observation the second line was very weak and its parameters could not be properly
constrained. In this case the line centroid was frozen at the outburst value of 
7.13\,keV. The width was again coupled with the Fe~K$_\alpha$ value. 
The residuals decreased from $\chi^2=721$ for 477 dofs
to the overall best fit values of $678$ for 474 dofs, including two CRSFs, when the
Fe~K$_\beta$ line was included and an F-test showed only  
a $2.8\times 10^{-6}$ probability that this improvement is of a statistical nature.\footnote{
See however \citet{Protassov:2002} about the usage of the F-test in line-like features. }
Another possibility for modeling the
observed residuals at $\sim 7$\,keV is the addition of an
Fe~K-edge, located at 7.1\,keV. When including the edge, the 
improvement is only marginal ($\chi^2$ of 712  for 474 dofs) 
and the best fit energy is $\sim 7.6$\,keV. Freezing the energy to the value of 7.1\,keV did not improve the fit. 
The remaining residuals indicated that an additional Fe~K$_\beta$ line
would still be necessary.  The measured energies of the Fe~K$_\alpha$
and K$_\beta$ lines were slightly higher than the expected laboratory
values, but still in the same order of magnitude of the energy scale
uncertainties of the instrument ( $\sim \pm$ 20 eV @ 6\,keV according
to the ABC guide).\footnote{An inflight energy calibration using the
Mn-calibration sources of the XIS detectors could not be performed, due to the 
usage of the 1/4 window mode, which excludes the corner 
regions where the calibration sources are located.}

Comparing the equivalent widths (EQWs) of the Fe~K$_\alpha$ and
K$_\beta$ lines in the two observations showed no significant change during the outburst. 
The Fe~K$_\alpha$ line showed an EQW of $
51\pm4$\,eV for both models, whereas the
Fe~K$_\beta$ line showed an EQW of $8\pm3$\,eV for the \texttt{cutoffpl}
model and $10\pm3$\,eV in the \texttt{compTT} model. 
For the second observation, the Fe~K$_\alpha$ EQWs were $46\pm10$\,eV 
and $49\pm10$\,eV, and the K$_\beta$ EQWs were $8\pm 2$\,eV and $11\pm 9$\,eV 
for the \texttt{cutoffpl} and \texttt{compTT}  models, respectively.  
The similarity of the Fe line EQW between both observations indicates that the 
source of the Fe lines is near to the source of the continuum (see discussion). 

 \subsection{Cyclotron resonance scattering feature} 
In the burst observation a strong residual at $\sim 50$\,keV
was observed in the HXD (Fig.~\ref{fig:on_spec}, b). 
 When including a CRSF-like feature with a Gaussian optical
depth profile (\texttt{gabs}) the best fit was obtained with a 
centroid energy of $58.2^{+0.8}_{-0.4}$\,keV, significantly improving the $\chi^2$ from
1728 with 480 dofs to 752 for 477 dofs for the \texttt{cutoffpl}
model, confirming the discovery of the CRSF in the \textsl{RXTE} data by \citet{Doroshenko:2010}.   
The width of the line was $14.1^{+3.5}_{-3.1}$\,keV and the
optical depth $\tau$ was $60.1^{+5.9}_{-1.5}$.  Using the
\texttt{compTT} model, similar values for $E_\text{cyc}$ and $\sigma$
were obtained for the outburst data. In this case the optical depth was found to be 
$23.6^{+10.4}_{-6.9}$. 

In the \texttt{cutoffpl} model, a second absorption line, using the \texttt{gabs} model, 
in the GSO data of the outburst spectrum at $\sim
110$\,keV improved the $\chi^2$ to 678 with 474
dofs (see Fig.~\ref{fig:on_spec}c). An F-test probability of $0.144$ indicates that
the line may not be significantly detected. However, an increase of the GSO Non X-ray Background
(NXB) by 1 or 2\% decreased the GSO counts to a non-detectable value. 
Another possible systematic effect is the decay of the $^{153}$Gd
instrumental line, which is activated when the satellite passes through
the south atlantic anomaly (SAA), and thus  creating a background line at
$\sim 150$\,keV. This could impose a deviation at $\sim 110$\,keV if not accurately represented 
in the modeled background. Therefore caution is advised when interpreting this feature. 

In the lower luminosity observation only an indication of the fundamental CRSF was observed
(see Fig.~\ref{fig:off_spec}). The line was fitted in 2 steps, where first the centroid energy and width were 
fixed to the outburst values, and the optical depth was left as a free parameter. 
With the addition of this CRSF, the \texttt{cutoffpl} model improved slightly from a 
$\chi^2$/dofs of 605 / 432 to 574 / 431 with an optical depth of $\tau = 15.4^{+0.64}_{-0.63}$.  
In the second step, the CRSF parameters were left free and the centroid energy decreased 
to $47.4^{+3.2}_{-2.3}$\,keV, whereas the optical depth decreased to $\tau=6.0^{+3.0}_{-2.1}$ and 
the width decreased to $\sigma = 5.7^{+2.0}_{-1.7}$\,keV.  With a $\chi^2$/dofs of 567 / 428  the best fit improved 
further  (see Tab. \ref{tab1} for details). 
To test the significance of the CRSF in the second observation, 100\,000 PIN spectra were Monte 
Carlo simulated using the null hypothesis approach where a spectrum, including Poisson noise, 
was created using the best fit parameters without an CRSF line.
The simulated spectra were then fitted with a continuum model and an additional \texttt{gabs} absorption 
line to test how often such a line could be detected in the spectral noise. The width was constrained to a 
value of $\sigma$ between 3.5\,keV and 8.5\,keV, so that neither very small features nor broad parts of the continuum were 
modeled. Out of the 100\,000 simulations, 42\% of the best fit results showed a non-zero value of $\tau$. 
When comparing the simulated $\tau$ distribution with the best fit result of $\tau = 6.0^{+3.0}_{-2.1}$,  $\sim 96$\% 
of all simulated fits showed a value between 0 and the measured value. When taking the errors of the 
measured data into account ($\tau<3.9$) the number is reduced to $\sim 90$\%.

For the \texttt{compTT} model, the addition of the fundamental CRSF with the frozen energy and width created 
residuals in the $10-20$\,keV range, which could be significantly reduced by decoupling the optical 
depth of the \texttt{compTT} model in the PIN data from the XIS data. The best fit optical depth of the CRSF 
line was $27.1^{+3.2}_{-3.0}$ changing the $\chi^2$ / dof values from 639 / 429 to 567 / 428, respectively. 
When leaving the CRSF parameters free, the width of the CRSF 
increased to $>30$\,keV, making it part of the continuum and tampering with the results. 
Freezing the CRSF parameters to the values obtained with the \texttt{cutoffpl} model did not result 
in a satisfactory overall fit. 

An alternative approach to describe the CRSF is to replace the
\texttt{gabs} model with the XSPEC model \texttt{cyclabs}
\citep{Mihara:1990}, described by the resonance energy $E_\text{res}$,
the resonance width $\sigma_\text{res}$ and the resonance
depth $\tau_\text{res}$. Using this model does not improve the fits 
significantly, and actually results in a slightly worse $\chi^2/$dofs value 
of 782 / 475 for the outburst data. Note that in the case of \texttt{cyclabs} 
the ratio between the energy of the fundamental and first harmonic line 
are fixed to 2, resulting in 1 more degree of freedom. 
The observed $E_\text{res}$ of $49.5^{+0.8}_{-0.8}$\,keV is of the order of $\sim 20$\% below 
the centroid energy obtained with the \texttt{gabs} model. This discrepancy stems from the use of a different  
calculation of the line energy and is of the order of $10-20$\,\% lower than the measured \texttt{gabs} energy 
\citep[see][ for details]{Nakajima:2010}. 
No significant changes in the continuum parameters have been observed when using the \texttt{cyclabs} model.
A width of $\sigma_\text{res} = 19.4^{+2.8}_{-1.3}$\,keV and a depth of $\tau_\text{res} = 1.45^{+0.10}_{-0.06}$
are obtained for the outburst observation. In the second observation the resonance energy shows a slight decrease 
($E_\text{res}=45.7^{+2.3}_{-2.0}$\,keV), and only slightly smaller than the \texttt{gabs} values. These results reflect again the difficulty of constraining the possible CRSF parameters  in the second observation. 
The width and depth,  $\sigma_\text{res} = 5.6^{+3.7}_{-2.4}$\,keV  and $\tau_\text{res} = 0.46^{+0.19}_{-0.14}$, are consistent with the 
\texttt{gabs} result showing a decrease in width and depth for both observations.

 \begin{figure}
 \plotone{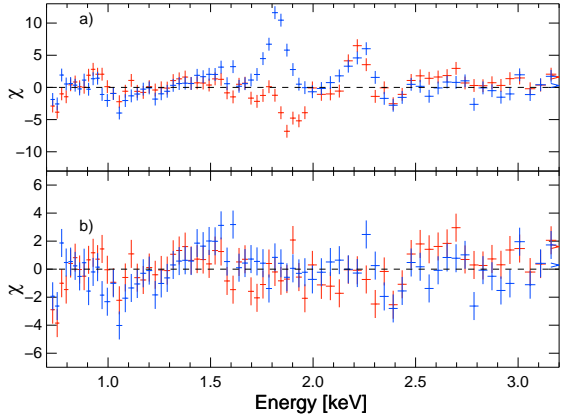}
\caption{Instrumental residuals at lower energies for XIS 1 (blue) and
  XIS 03 (red) with the \texttt{cutoffpl} model. a) shows the clearly apparent lines are the Au~K$_\alpha$ line at $\sim
  2.1$\,keV and the Si~K$_\alpha$ line at $\sim 1.8$\,keV. Additionally
  a possible Ni~K$_\alpha$ line at $\sim 0.9$ which improves the fit slightly. b) shows the best fit residuals when the lines are included.}
\label{fig:lowenergy}
\end{figure}

\subsection{Low energy calibration issues} \label{sec:lowen}
Strong residuals at lower energies, i.e. between 1.5 and 2.5 keV (see
Fig. \ref{fig:lowenergy}) were observed in both data sets.  
A comparison with known background properties of the
\textsl{Suzaku}/XIS instrument \citep{Yamaguchi:2006} showed that
these features are identical to the known instrumental Si K$_\alpha$
and Au K$_\alpha$ lines, located at 1.74\,keV and 2.12\,keV,
respectively.  In many observations of bright sources, e.g. 4U\,1907+07
\citep{Rivers:2010} and LMC\,X$-$3 \citep{Kubota:2010}, these energy
bands are explicitly excluded.  In this case, modeling 
these lines with two Gaussian emission features was 
sufficient to minimize the residuals and improve the $\chi^2$/dofs
from 1295 / 483 (without the lines) to the best fit value of 678 / 474. Note that the Au
K$_\alpha$ line at $\sim 2.1$\,keV could be described by the same
Gaussian emission line at $2.21\pm0.01$\,keV for both FI and BI XIS
instruments, whereas the Si K$_\alpha$ line appears as an emission line
for the BI XIS1 at $1.82\pm0.01$\,keV (Fig.~\ref{fig:lowenergy}, blue)
and a negative Gaussian line at $1.89\pm0.01$\,keV for the FI XIS 03
combination (Fig.~\ref{fig:lowenergy}, red), indicating that for the
FI instrument the line is either over-subtracted or not properly
energy calibrated, leading to a dip in the spectrum. By using two
independent lines at 1.89\,keV (FI) and 1.82\,keV (BI) the residuals
can be well described. An additional Gaussian component at $0.93^{+0.01}_{-0.02}$\,keV, 
located close to the Ni K edge at 0.897\,keV, slightly improves the residuals 
($\chi^2$/dofs of 731 / 477 without the line, compared to the best-fit value of 678 / 474) 
This improvement indicates a possible systematic error of the calibration at lower energies
for bright sources. This line is not very pronounced and can be
omitted when using the \texttt{compTT} model.

 \begin{figure}
\plotone{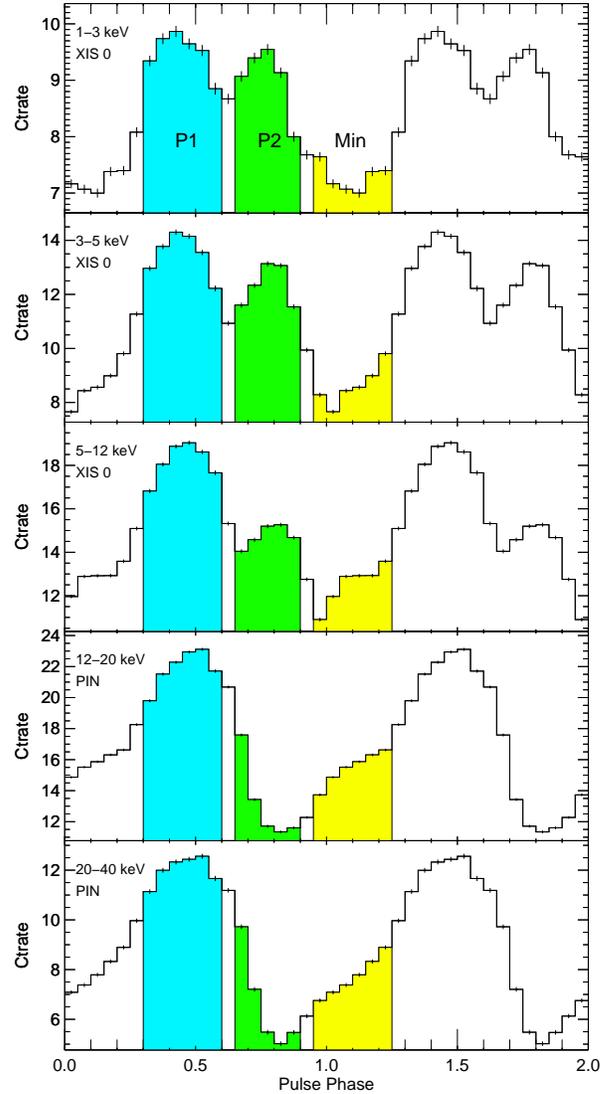}
\caption{XIS and PIN  pulse profiles for different energy bands for the outburst observation. P1, P2 and MIN indicate the 
regions used for the phase resolved spectral analysis.}
\label{fig:profile1}
\end{figure}

\begin{figure}
 \plotone{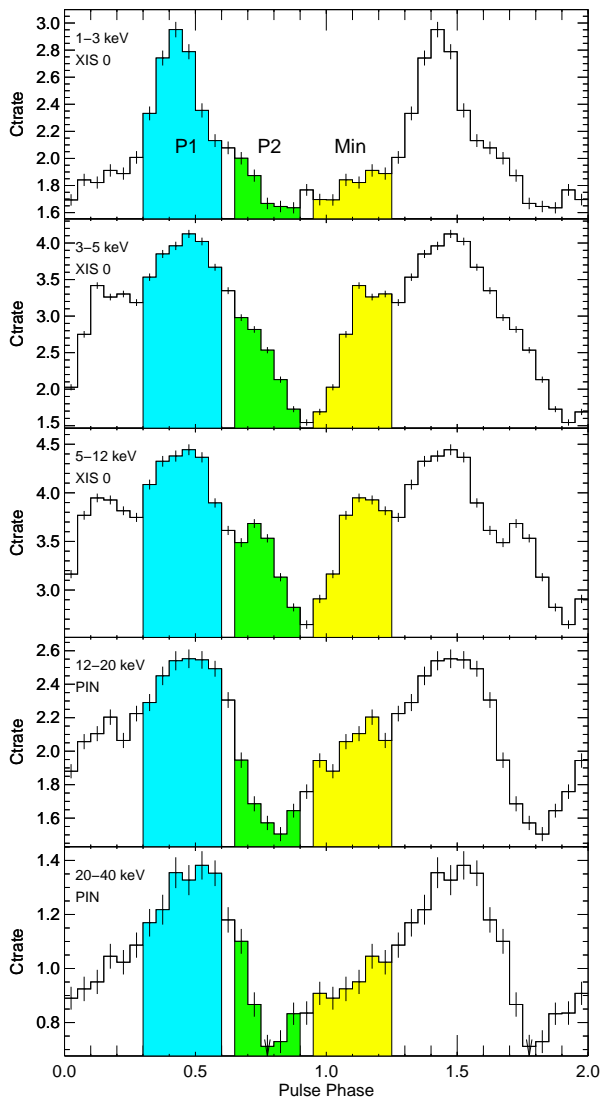}
\caption{Same as Fig.~\ref{fig:profile1} for the second observation. The highlighted regions indicate the same phase bins as in the outburst observation for a direct comparison}
\label{fig:profile2}
\end{figure}

\section{Phase resolved analysis} 
For a phase resolved analysis the XIS and PIN data of both
observations were folded with the \textsl{RXTE} determined pulse
period from \citet{Doroshenko:2010} of $P_\text{spin} = 407.719$\,s,
$\dot{P}_\text{spin} = -4.6\times10^{-7}$\,s/s and the MJD epoch of
54841.62.  Pulse profiles with 20 phase bins were created for five
different energy bands in the $1-10$\,keV energy range for XIS and
$12-40$\,keV energy range for PIN for both observations
(Fig.~\ref{fig:profile1} and Fig.~\ref{fig:profile2}). The statistical quality of the
GSO data precluded the creation of pulse profiles and further spectral analysis. 

During the outburst, the pulse profile consisted of two peaks, where the
main peak (P1, pulse phase $0.3-0.6$) stayed dominant throughout all
energy bands and became broader towards higher energies
(Fig.~\ref{fig:profile1}). The second peak (P2, pulse phase $0.65-0.9$) disappeared at energies
above $\sim 10$\,keV, consistent with the \textsl{RXTE} observations
by \citet{Doroshenko:2010}.  The third region in
Fig.~\ref{fig:profile1} indicates the minimum of the pulse profile
(MIN, pulse phase $0.95-1.25$), as determined from the lower energies of the XIS instrument.

In comparison, the pulse profile of the second observation
(Fig.~\ref{fig:profile2}) showed a similar Peak P1 throughout the whole
energy band, although narrower at lower energies. No second peak was observed 
at the position of P2, although a small ``bump'' in the $5-12$\,keV energy range was still visible. At lower energies, i.e. for
the $3-5$ and $5-12$\,keV energy band, a small peak showed up on the
opposite site of P1, the pulse phase where the minimum was defined
in the outburst data. 

\subsection{Phase resolved spectroscopy}
Spectra were extracted for three different pulse phases throughout the
outburst: P1, P2 and MIN. 
XIS and PIN spectra were used and the same XIS grouping as in the
phase averaged data was applied. The PIN data was again grouped to
include at least 100 counts per spectral bin. The same spectral model was used 
as in the phase averaged analysis and the best-fit results are summarized in Table~\ref{tab2}. 
In all pulse phases the CRSF
component was visible in the spectra.  Throughout the pulse phase 
the best-fit values for the galactic and intrinsic
$N_\text{H}$ values did not change significantly. 
Note that in both models a slight decrease of the covering fraction
during P2 can be observed.

\begin{deluxetable*}{lccclccc}
	\tablewidth{0pc}
	\tabletypesize{\scriptsize}
	\tablecaption{Phase resolved outburst spectral parameters. Same models as in Table 1\label{tab2}}
	\tablecolumns{8}
\startdata
\hline
\hline
\multicolumn{1}{c}{ Parameter}  &  \multicolumn{3}{c}{Cutoffpl} & \multicolumn{1}{c}{Parameter}  &  \multicolumn{3}{c}{CompTT}  \\                 
                            							&  Peak 1         			& Peak 2        			   &     Minimum  & & Peak 1         & Peak 2        &     Minimum        \\
\hline                            
\text{phabs} $N_\text{H} [10^{22}$/cm$^2$] 	&$ 1.35^{+0.02}_{-0.03}$	 &  $1.38^{+0.03}_{-0.01}	$ &$ 1.21^{+0.04}_{-0.04}$	& 				&  $1.28^{+0.04}_{-0.05}	$  &$ 1.22^{+0.6}_{-0.5}$	 &  $1.20^{+0.04}_{-0.05}	$        \\
\text{pcfabs} $N_\text{H} [10^{22}$/cm$^2$] 	&$ 11.1^{+0.8}_{-0.5}$	 &  $10.6^{+0.4}_{-0.2}	$ &$ 11.7^{+1.1}_{-1.1}$		& 				&  $11.6^{+0.8}_{-0.8}	$  &$ 11.0^{+0.9}_{-1.0}$	 &  $12.16^{+1.09}_{-0.89}	$        \\
covering fract.	                                                        &$ 0.66^{+0.03}_{-0.03}$	 &  $0.52^{+0.1}_{-0.02}	$ &$ 0.67^{+0.05}_{-0.05}$	& 				&  $0.70^{+0.03}_{-0.04}	$  &$ 0.58^{+0.6}_{-0.7}$	 &  $0.70^{+0.04}_{-0.05}	$        \\ 
blackbody kT [keV]     					&$ 0.52^{+0.03}_{-0.02}$	 &  $0.58^{+0.01}_{-0.02}	$ &$ 0.50^{+0.02}_{-0.02}$	& 
	&  $0.54^{+0.03}_{-0.02}	$  &$ 0.59^{+0.03}_{-0.03}$	 &  $0.52^{+0.03}_{-0.02}	$        \\ 
blackbody norm   $[10^{-3}]^\dagger$				&$ 7.0^{+1.0}_{-1.6}$	 &  $2.3^{+0.6}_{-0.4}	$ &$ 8.1^{+1.7}_{-0.7}$		& 
	&  $12.3^{+1.7}_{-1.9}	$  &$ 8.9^{+1.6}_{-1.6}$	 &  $10.1^{+1.6}_{-1.4}	$        \\
$E_\text{cut}$ [keV]     					&$ 29.5^{+7.2}_{-3.2}$	 &  $18.3^{+0.8}_{-0.4}$ &$ 10.7^{+1.3}_{-0.9}$		& compTT T$_0$ [keV]
	&  $1.31^{+0.10}_{-0.07}	$  &$ 1.41^{+0.04}_{-0.05}$	 &  $1.48^{+0.11}_{-0.07}	$        \\ 
$\Gamma$               						&$ 0.42^{+0.06}_{-0.04}$	 &  $0.56^{+0.04}_{-0.02}	$ &$ -0.13^{+0.09}_{-0.08}$	& compTT kT [keV]
	&  $7.67^{+1.39}_{-0.38}	$  &$ 11.91^{+9.06}_{-2.77}$	 &  $6.79^{+0.75}_{-0.21}	$        \\ 
$A_\text{pl}^\dagger$							&$ 0.12^{+0.01}_{-0.01}$	 &  $0.14^{+0.01}_{-0.00}	$ &$ 0.04^{+0.004}_{-0.004}$	& compTT $\tau_\text{p}$
	&  $6.31^{+0.31}_{-0.40}	$  &$ 4.00^{+0.47}_{-0.45}$	 &  $6.60^{+0.38}_{-0.62}	$        \\
&&&	& compTT norm $^\dagger$&  $0.11^{+0.01}_{-0.01}	$  &$ 0.05^{+0.02}_{-0.01}$	 &  $0.08^{+0.01}_{-0.01}	$        \\
$E_\text{CRSF}$ [keV] 					&$ 66.6^{+0.9}_{-2.3}$	 &  $55.5^{+0.5}_{-0.7}	$ &$ 52.8^{+2.9}_{-2.0}$		& 
	&  $57.9^{+8.7}_{-4.2}	$  &$ 57.7^{+2.0}_{-4.1}$	 &  $51.0^{+4.9}_{-2.4}	$        \\ 
$\sigma_\text{CRSF}$ [keV]				&$ 19.2^{+3.6}_{-2.7}$	 &  $11.4^{+0.3}_{-0.2}	$ &$ 9.3^{+2.5}_{-1.6}$		& 
	&  $13.5^{+15.5}_{-3.4}	$  &$ 14.9^{+2.9}_{-3.1}$	 &  $9.0^{+4.6}_{-2.0}	$        \\
$\tau_\text{CRSF}$ 					 	&$ 130.8^{+7.0}_{-19.6}$	 &  $44.1^{+2.3}_{-3.3}	$ &$ 22.7^{+14.6}_{-7.3}$		& 
	&  $25.6^{+45.4}_{-12.3}	$  &$ 79.5^{+64.8}_{-40.2}$	 &  $13.5^{+20.4}_{-4.7}	$        \\
$E_{\text{Fe~K} \alpha}$ [keV] 		&$ 6.41^{+0.01}_{-0.01}$	 &  $6.41^{+0.01}_{-0.01}	$ &$ 6.42^{+0.01}_{-0.01}$	& 
	&  $6.41^{+0.01}_{-0.01}	$  &$ 6.41^{+0.01}_{-0.01}$	 &  $6.42^{+0.01}_{-0.01}	$        \\ 
$E_{\text{Fe~K} \beta}$ [keV]		&$ 7.16^{+0.05}_{-0.05}$	 &  $7.08^{+0.06}_{-0.06}	$ &$ 7.09^{+014}_{-0.11}$	& 
	&  $7.16^{+0.04}_{-0.05}	$  &$ 7.09^{+0.06}_{-0.05}$	 &  $7.12^{+0.14}_{-0.12}	$        \\ 
Eq. Width K$_\alpha$ / K$_\beta$ [eV]  		&$ 37^{+9}_{-6}/7^{+7}_{-7}$	 		 &  $59^{+8}_{-9}/11^{+9}_{-9}	$ 		   &$ 64^{+9}_{-9}/8^{+7}_{-8}$				& 
	&  $39^{+5}_{-6}/10^{+7}_{-7}	$  &$58^{+7}_{-4}/13^{+7}_{-8}$	 &  $62^{+9}_{-8}/8^{+9}_{-8}	$        \\ 
	
Flux$_{2-10\,\text{keV}}   [10^{36}$ ergs/sec ]$^\ddagger$            &   10.5 & 8.14 & 7.5            &  & &\\
C$_\text{XIS1}$/C$_\text{PIN}$ 			& 0.98 / 1.1 			&  0.97 / 1.04   			&  0.99/1.16 	 			&
	& 1.04 / 1.23    &  0.99 / 1.19 &   1.08 / 1.23 \\
$\chi^2$/dofs     				&$ 498 / 431$			&$ 508 / 420$ 			&   474 / 427 	 			&
	& 528 / 438     &   492 / 423  &   542 / 430\\
\hline
\enddata
\tablecomments{$\dagger$ Units in Photons keV$^{-1}$ cm$^{-2}$ s$^{-1}$, $\ddagger$ unabsorbed flux using a distance of 5\,kpc  }
\end{deluxetable*}

The most significant variation in the \texttt{cutoffpl} model was that of
the cutoff energy $E_\text{cut}$ which decreased from $29.5^{+7.2}_{-3.2}$\,keV in P1,
to $18.3^{+0.8}_{-0.4}$\,keV in P2 and down to $10.7^{+1.3}_{-0.9}$\,keV in MIN. At the same time the power
law index $\Gamma$ varied from $0.42^{+0.06}_{-0.04}$ (P1), 
to $0.56^{+0.04}_{-0.02}$ (P2), and to 
$-0.13^{+0.09}_{-0.08}$ for MIN. In the three
phases the CRSF  centroid energy changed from
$66.6^{+0.9}_{-2.3}$\,keV in P1, and declined to
$55.5^{+0.5}_{-0.7}$\,keV throughout P2, and $52.8^{+2.9}_{-2.0}$\,keV
for MIN. The CRSF width declined from $19.2^{+3.6}_{-2.7}$\,keV in P1 to
$11.4^{+0.3}_{-0.2}$\,keV in P2, and to $9.3^{+2.5}_{-1.6}$ \,keV in MIN.

For the \texttt{compTT} model most parameters did not change throughout
the pulse profile. The observed increase in the plasma temperature and the decrease of the 
optical depth $\tau_\text{p}$ is a known systematic anti-correlation (see discussion). 
As with the phase averaged data, the CRSF did not show the same changes throughout the 
pulse profile, as in the \texttt{cutoffpl} model. The best-fit values for the CRSF centroid energy were
$57.9^{+8.7}_{-4.2}$\,keV for P1, $57.7^{+2.0}_{-4.1}$\,keV for P2, and
$51.0^{+4.9}_{-2.4}$\,keV for MIN, and the widths were
$13.5^{+15.5}_{-3.4}$\,keV,$14.9^{+2.9}_{-3.1}$\,keV, and $9.0^{+4.6}_{-2.0}$\,keV
for P1, P2 and MIN, respectively.  Fe~K$_\alpha$ and Fe~K$_\beta$
energies were consistent with $\sim6.4$\,keV and $\sim7.1$\,keV, similar to the \texttt{cutoffpl} values.

\section{Discussion}
This paper presents an analysis of the two \textsl{Suzaku} observations of the
Be/X-Ray binary 1A\,1118$-$61 during the peak of its outburst in 2009,
January and $\sim 13$ days later. A CRSF, detected with \textsl{RXTE}
at $\sim 55$\,keV, could be observed in both observations, although
the significance is lower in the second observation. 
An Fe K$_\beta$ line at $7.13$\,keV has been observed in addition to the strong, narrow
 Fe K$_\alpha$ line at 6.4\,keV.  The broad band continuum was
modeled with the empirical \texttt{cutoffpl} model, including an
additional 10\,keV systematic component to improve the residuals. Softening of
the power law index $\Gamma$ between peak and decay had been observed
in an earlier outburst and could be confirmed.  The Comptonization model
\texttt{compTT}, where the 10\,keV component was not needed, has also been applied. 
The pulse profiles at lower energies changed from a two peaked to a single 
peaked profile between both observations. Phase resolved spectral analysis 
was performed for both observations and the same models as in the phase 
averaged analysis were applied. 

\subsection{Outburst behavior}
The third observed outburst of 1A\,1118$-$61 follows a pattern similar to 
the second outburst from 1992, i.e. a strong peak lasting $\sim 3$
weeks and an elevated level of emission up to 6 weeks afterwards. The
time between outbursts was in both cases $\sim 6200$\,days, corresponding 
to  $\sim 17$ years, indicating that the outburst behavior could 
be periodic on very long time scales. 
The proposed orbital periods are $\sim 350$\,days \citep{Corbet:1986}, $\sim 58$\,days \citep{Reig:1997}. 
and most recently 24\,days established by \citet{Staubert:2010} using the delay in pulse arrival time of \textsl{RXTE} monitoring observations throughout this outburst. Using the latter method the  
\textsl{Suzaku} light curves are in full agreement with the \textsl{RXTE} results (Staubert, priv. comm). 
A period of 24\,days would put 1A\,1118$-61$ in the wind accretor region on the 
``Corbet'' diagram \citep{Corbet:1986}, making it a very unique source for a Be system. 

A very similar scenario was introduced by \citet{Villada:1999}, who
monitored the $H_\alpha$/$H_\beta$ emission before and after the
second outburst and proposed that the optical companion, Hen~$3-640$, has an extended large
envelope where a weak interaction with the NS can 
occur. 
In the scenario of \citet{Villada:1999}, the NS is orbiting the O star in an environment with gradually 
increasing density, until a steady accretion disk is created and
the X-ray flux suddenly increases. The sudden increase in the accretion 
material would provide the torque on the NS to produce the observed changes in the pulse period. 
The surrounding material is then swept out in a short time and the system returns to quiescence. 
According to \citet{Villada:1999}, an interval of 17 years between the outbursts is a reasonable time scale 
to accumulate enough material between the outbursts. 

\subsection{Cyclotron features} 
In the outburst, a CRSF has been observed at $\sim 55$\,keV 
and 
the parameters are consistent with the \textsl{RXTE} data \citep{Doroshenko:2010}. 
A CRSF in the second observation is expected but can not be confirmed with sufficient significance in the 
established data. 

CRSFs are generated by electrons
with energies close to the ones determined by the discrete 
energy states of the Landau levels being excited to higher levels followed by 
 de-excitation emitting a photon. 
In this process,
photons from the accretion column above the magnetic poles with a
resonant cyclotron line energy, are resonantly scattered out of the
line of sight creating an absorption line-like feature. These features 
provide a direct measure of the magnetic field strength close to the NS
surface, where the fundamental cyclotron energy is given by
\begin{equation} 
E_{\text{cyc}} = \frac{\hbar e B}{m_e c} \frac{1}{1+z} = \frac{11.6
  \text{\,keV}}{1+z}\times B_{12}, 
\end{equation}
where $B_{12}$ is the magnetic field strength near the NS
surface in units of $10^{12}$\,Gauss and $z$ is the gravitational
redshift. Assuming a typical NS mass of 1.4\,\Msun and
NS radius of 10\,km gives $z = 0.3$. For 1A\,1118$-$61 the
CRSF at $\sim$~55\,keV corresponds to a magnetic field of $\sim
6.4\times 10^{12}$\,Gauss, making it, together with the $\sim 50$\,keV 
line of A\,0535+26, one of the strongest observed magnetic fields on an accreting neutron star in 
a binary system.

A possible feature in the residuals at $\sim 110$\,keV could indicate the
existence of a first harmonic line, as observed in multiple sources,
such as 4U\,0115+63 \citep{Heindl:1999}, 4U\,1907+09 \citep{Cusumano:1998,Makishima:1999}
and Vela\,X$-$1 \citep{Makishima:1999,Kreykenbohm:2002}. Including a
\texttt{gabs} line at this energy does not significantly improve the fit, though a $^{153}$Gd
instrumental line at $\sim 150$\,keV, which is due to electron-capture \citep{Kokubun:2007}, 
as well as possible systematic uncertainties in the
background weaken the significance of the detection further. 

Assuming that the observed width of the CRSF is due to the 
Doppler broadening of the electrons responsible for the resonances, one can 
estimate the inferred plasma temperature in the CRSF region using equation 4.1.28 from \citet{Meszaros:1992}:
\begin{equation}
\Delta \omega_\text{D} = \omega_{\text{Cyc}} \left( \frac{2 kT}{m_e c^2} \right)^{1/2} |{\text{cos}(\Theta})|, 
\end{equation} 
where $\Delta \omega_\text{D}$ corresponds to the CRSF FWHM which is calculated from $\sigma_\text{Cyc}$, $\omega_{\text{Cyc}}$ is
equivalent to the CRSF centroid energy $E_\text{Cyc}$ and $\Theta$ is the angle between magnetic field lines and the line of sight. 
With $\text{cos}(\Theta) =1$ the lower limits for the plasma temperature are $kT \approx 20.8$\,keV for the \texttt{cutoffpl} model and $kT \approx 12.6$\,keV for the \texttt{compTT} model when using the 
values from Tab. 1. 

Although the existence of a CRSF in the second observation is only marginal, 
it is very likely that such a line exists at lower luminosities.
Luminosity dependance of CRSFs have been observed in multiple other binary system.  
In Be-Binaries, such as V\,0332+53, 4U\,0115+63, and X\,0331+53, where \citet{Tsygankov:2006} and \citet[][2010]{Nakajima:2006} found a negative 
correlation between the CRSF centroid energy and the luminosity of the source for luminosities approaching the Eddington luminosity. 
\citet{Staubert:2007} on the other hand found a positive CRSF$-$luminosity correlation in the low mass X-ray binary Her\,X$-$1 for luminosities far below the Eddington luminosity. 

The type of correlation seems to depend on whether the observed luminosity is above or below the 
Eddington luminosity. When the luminosity is above the Eddington luminosity, the infalling protons
start to interact before they are part of the accretion column and a
``shock front'' region, where the CRSF most likely occurs, is created. 
With increasing luminosity, the proton interaction 
occurs farther away from the NS surface, where the magnetic field is 
lower and therefore the observed CRSF is seen at lower energies.
Below the Eddington luminosity, the proton interaction does not 
occur above the accretion column but is part of it. When the 
luminosity increases, the accretion ``pressure'' increases as well 
and the region where the CRSF is created is pressed closer to the 
NS surface, where the magnetic field is higher. This results in a positive 
CRSF-luminosity correlation, as observed in Her X$-$1.  

The flux levels (Tab.~1) obtained for both observations showed that the 
first observation is most likely above, and the second observation is definitely 
below the Eddington luminosity of the system, so that an anti-correlation between CRSF centroid energy and luminosity are expected.

\subsection{Continuum comparison} 
When comparing both observations, changes in the broad band continuum parameters 
were observed. 
During the 1974 outburst
\citet{Ives:1975} observed a harder spectrum in the peak compared to  
later observation.  Looking at the
\texttt{cutoffpl} model, the \textsl{Suzaku} data showed a similar behavior with a
very hard spectrum (power law index $\Gamma=0.35$) at the luminosity peak,
and a much softer power law index of $\Gamma \sim 1$ in the second observation. Note
also that the cutoff/folding energy increased with declining luminosity. A
similar behavior has been observed in a number of different HMXB
transients, such as A\,0535+26 \citep{Caballero:2009} and V\,0332+53 \citep{Mowlavi:2006}.  
This is in contrast to EXO\,2030+375 \citep{Reynolds:1993}, where
the observed folding energy increased with lower luminosity. 
\citet{Soong:1990} interpreted this parameter in phase resolved results and concluded that 
the folding energy reflects a change of the viewing angle on the accretion column, 
which allows a deeper look into the emission region and therefore directly correlates with the 
observed electron plasma temperature. 
The change observed in the phase averaged observations follows a similar 
reasoning, and at lower luminosities the emission region is closer to the NS, where the 
observed plasma temperatures are expected to be higher. 

A comparison of \texttt{compTT} parameters between
\textsl{Suzaku} and \textsl{RXTE} \citep{Doroshenko:2010} showed
consistent results for the spectral parameters in the
outburst. In contrast to the \textsl{RXTE} results, the observed
column density $N_\text{H}$ is higher and consistent with the results
obtained by the \texttt{cutoffpl} model. The main difference between these results and 
the \textsl{RXTE} data lies in the use of a combined Galactic and intrinsic column density and 
the additional need of a low energy black body, so that these values cannot be compared directly. 
Note that the \textsl{RXTE}/PCA instrument used in \citet{Doroshenko:2010} is not sensitive to 
data below 3\,keV and therefore a partial covering, as well as a black body component could not be modeled. 
Both, the black body temperature for the low energy excess and the photon 
seed temperature in the \texttt{compTT} model decreased by $\sim 1/3$ between 
outburst and the second observation, indicating a correlation between both parameters. 
A possible explanation is that the source of the \texttt{compTT} seed photons is the same as the soft 
excess modeled by the black body component. The observed change in the optical thickness indicates 
the the observed material is optically thick in the outburst and gets optically thinner for the second observation. 

Although an increase in the plasma temperature of the \texttt{compTT} model was 
observed after the outburst, i.e., mainly consistent with the interpretation of the 
\texttt{cutoffpl} parameters, one has to be careful with a direct interpretation of the spectral parameter. 
The optical depth $\tau$ and the electron plasma temperature $kT_\text{e}$ show a 
very strong negative correlation \citep[e.g.][]{Wilms:2006}, and the best-fit values cannot be 
used directly for interpretation. Following the definition in \citet{Rybicki:1979}, the Compton $y$ parameter: 
\begin{equation} 
y = \frac{4kT_\text{e}}{m_\text{e}c^2} \text{max}(\tau, \tau^2), 
\end{equation} 
can help in the physical interpretation. 
\citet{Reynolds:2003} used the Compton $y$ parameter in the description of accretion disk coronae around 
black holes. 
A value of $y \approx1$ or slightly higher means that the average emitted photon 
energy increases by an ``amplification factor'', $A(y) \approx \text{exp}(y)$, and is referred 
to as ``unsaturated inverse Comptonization''. For  $y\gg1$ the average photon 
energy reaches the thermal energy of the electrons. This case is called 
the ``saturated inverse Comptonization''. 
Using $kT$ and $\tau_p$ results in $y=2.13\pm0.2$ for the outburst data and  
$y=1^{+9}_{-0.7}$ for the second observation. Typical calculated values of $y$-parameters are 
smaller than 1, e.g., $\sim 0.5$ for Cyg\,X$-$1 \citep{Reynolds:2003}, 
$\sim 0.2$ for 4U\,2206+54 \citep{Torrejon:2004}, and $\sim 0.6$ for XV\,1832$-$330 
\citep{Parmar:2001}. Note that the second observation is very badly constrained 
due to the large error bars on the electron plasma temperature and the 
optical depth and therefore cannot be used for interpretation. The 
$y$ parameter for the \textsl{Suzaku} outburst indicates that the system is in, 
or very close, to a ``saturated inverse Comptonization'' state. 
 
\subsection{Fe lines} 
In both observations an  Fe\,K$_\alpha$   and  Fe\,K$_\beta$ emission line has been observed, 
where the Fe\,K$_\alpha - $ Fe\,K$_\beta$ normalization ratio of 12\% is consistent with neutral material. 
The Fe\,K$_\alpha$  EQWs were $51\pm4$\,eV and $46\pm10$\,eV for the first and second observation, 
respectively, while at the same time the $5-7$\,keV power law flux dropped from $6.50\pm0.01 \times 10^{-10}$ ergs cm$^{-2}$ s$^{-1}$ 
to $1.23\pm0.01\pm \times 10^{-10}$ ergs cm$^{-2}$ s$^{-1}$. The relatively constant EQWs imply that the Fe line 
emitting region is relatively close to the source of ionizing flux, so that the observed Fe line intensity adapts  
quickly to the changing incident flux. The Fe line normalization was $1.79 \pm 0.06\times10^{-3}$ photons cm$^{-2}$ s$^{-1}$
 and $3.15 \pm 0.3 \times10^{-4}$ photons cm$^{-2}$ s$^{-1}$ for the respective observations. The ratios between Fe normalization and 
 power law flux were consistent in both observations, $2.75\pm0.10$ and $2.56 \pm 0.27 \times 10^{6}$ photons/ergs, reflecting the two calculated EQWs.    
Using the continuum flux difference and the time between observations, the calculated change 
in power law flux was $\sim 0.415\pm0.001\times10^{-10}$ ergs cm$^{-2}$ s$^{-1}$ per day in the 12.7\,day period between the observations. 
A linear decrease could be assumed using the BAT data of Fig.~1. For the second observation, the Fe normalization was manually increased in XSPEC until the resulting EQW matched the upper limit of the measured EQW. 
This value corresponds to an Fe normalization of $\sim 3.9 \times 10^{-4}$ photons cm$^{-2}$ s$^{-1}$, assuming the measured continuum given above for the second observation. Together with the previously calculated constant of $2.56 \times 10^{6}$, a matching power law flux of $\sim 1.52$ could be calculated. Using this flux difference, together with the rate of change in the $5-7$\,keV energy range, one can then determine a maximum time delay which would still preserve the observed EQW within errors. 
The calculated upper limit to the delay between the Fe and X-ray emission region is $\sim 0.7$\,days. 
Note that this estimation is very simple and uncertainties from the orbital motion or from 
line of sight assumptions are not taken into account here. 

\subsection{Phase resolved description}
1A\,1118$-$61 shows a similar energy and luminosity pulse profile dependency 
as many other sources, e.g. 4U\,0115+63 \citep{Tsygankov:2007}, V\,0332+53
\citep{Tsygankov:2006} and A\,0535+26  \citep{Caballero:2009}. 
In the outburst observation, the observed
broad double peaked structure at lower energies changed to a single peak
profile above 10\,keV, where the main peak (P1)
broadened slightly with increased energy and the secondary peak
(P2) weakened. For the second observation, the pulse profile 
changed significantly. There was still a dominant primary peak at P1, although it is narrower in the 
lowest energy band.  The secondary peak is only marginally indicated
between $5-12$\,keV (see Fig. \ref{fig:profile2}) and absent in the other energy bands.  
On the other hand, between $3-12$\,keV a small peak or extended shoulder could be observed 
preceding the primary peak. 

To confirm the disappearance of the second peak, pulse fractions 
of the total counts of the P1/P2 regions were calculated for the indicated energy bands. 
The pulse fraction was defined as  (P1$_\text{countrate}-$P2$_\text{countrate}$) / P2$_\text{countrate}$
where the P1 and P2 regions are indicated in Fig. \ref{fig:profile1} and \ref{fig:profile2}. 
Table~\ref{tab:pf} shows that the pulse fraction increases towards higher 
energies in the outburst data, typical for a weakening of the second pulse. In the second observation the values do not vary  
for the different energy bands, and the small peak in the $5-12$\,keV band 
shows only a marginal smaller pulse fraction than in the other energy bands. 
\begin{deluxetable}{lcc}
	\tablewidth{0pc}
	\tablecaption{Pulse fractions for 1A\,1118$-$61 \label{tab:pf}}
\startdata
\hline
\hline
 Energy [keV] & Outburst  & 2nd Obs \\   
\hline
                    $1-3$        & 0.26 &  0.73  \\ 
                    $3-5$        &  0.31 &  0.90  \\ 
                    $5-12$      &  0.48 &  0.53  \\ 
                    $12-20$    &  1.00 &  0.78  \\ 
                    $20-40$   &  1.19  &  0.83  \\
\hline                    
\enddata
\end{deluxetable}

 \citet{Tsygankov:2007} described similar profiles and proposed
that a misalignment of the rotational and magnetic axes of the neutron
star leads to the case where one accretion column is observed whole,
whereas the second accretion column is partially screened by the
NS surface so that only the softer photons, which are
created in the higher regions of the accretion column, are observed.
When the overall luminosity, and therefore accretion rate, decreases,
the column height decreases and even the soft photons 
from the second pole are shielded. This behavior is in contrast to the 
variation observed in the cyclotron line production region, where the 
line forming region is closer to the NS surface at higher luminosities. 

For a more physical picture gravitational effects, such as light
bending, have to be taken into account when 
discussing pulse profiles \citep{Kraus:1995, Meszaros:1985}. 
For a canonical NS with a mass of 1.4\Msun and
a radius of 10\,km the visible surface of a NS is 83\%.  
With this increased surface visibility
the pulsed flux from both hot spots is visible over a longer part of
the pulse phase, and parts of the accretion column of the second peak
are still visible, although that hot spot is on the far side of the
NS. With decreasing luminosity, the hot spot size decreases and the 
visible fraction of the second accretion column disappears.

Pulse profile decomposition methods have been developed by
\citet{Kraus:1995}, and have
been applied to multiple sources, e.g. EXO\,2030+375 \citep{Sasaki:2010}
and A\,0535+26 \citep{Caballero:2011}. Under the assumption of a slightly
distorted magnetic dipole, with this method it is possible to
disentangle the contribution of the two emission regions, and
constraints on the geometry of the pulsar and on its beam pattern
can be obtained. The application of this method on the 1A1118
data presented here will be left for future work.

A more physical approach for a broad band spectrum has been introduced 
by \citet{Becker:2007}, using bulk and dynamical Comptonization of photons 
in the accretion column. First results on 4U\,0115+63 have been promising 
\citep{Ferrigno:2009} and 1A\,1118$-$61 is a good candidate for future tests.
\citet{Schoenherr:2007} developed a new physical model for the CRSFs based on 
Monte Carlo  simulations for the Green's functions for the radiative transport through a 
homogeneous plasma, which had been successfully applied for different sources, e.g., 
Cen\,X-3 \citep{Suchy:2008}. Due to advances in the overall code \citep{Schwarm:2010} the 
test of this model on this data set is beyond the scope of this paper. 

\section{Summary and Conclusions} 
In this paper we analyzed the broad band spectrum and pulse profiles of 1A\,1118$-$61
during the peak of its third observed outburst and compared the results with a second observation 
which occurred $\sim 2$\,weeks after the main peak.
The time between outbursts is consistent with a continuous low level accretion 
mechanism as suggested by \citet{Villada:1999}. 
A CRSF during the outburst has been confirmed at $\sim 55$\,keV indicating 
one of the highest known B-fields observed in HMXBs. 
In the second observation there is only a weak indication of a CRSF fundamental line. A change in the 
CRSF centroid energy with respect to luminosity would be expected and would help to understand the physical environment close to the NS surface.
Variations of the CRSF can also be observed in the phase resolved spectroscopy of the outburst observation. 
The calculated $y$ parameters show that the inverse Comptonization during the outburst is very close to saturation. 
The ratio between the Fe\,K$_\alpha$ and Fe\,K$_\beta$ normalization of $\sim 12\%$ 
shows that the emitting material is mostly neutral. 
Using the \texttt{compTT} model, one can deduce that the emitting material is optically thick during the outburst and optically thin 
in the second observation, although the known $kT - \tau$ correlation has to be considered. 
The pulse profile change for different energy bands and luminosities is similar to other observed HMXBs and 
can be explained with a misalignment of the rotation axis to the magnetic field axis. 
A change in the pulse profile shape with luminosity has been observed and indicates that the visibility of the second hot spot 
changes between the observations. 
Phase resolved analysis throughout the outburst also indicates a change in the observed magnetic field, 
which could be caused by different viewing angles onto the accretion column. Future work with a pulse deconvolution technique, as well as a more physical model will provide a better understanding of the involved physical processes.  

\begin{acknowledgements}
We thank the anonymous referee for his comments which improved the quality of this paper significantly. 
This work was supported by NASA grant NNX09AO91G for Suzaku's cycle 3. 
S. Suchy is funded by NASA grant NNX08AD72G and R. Rothschild is funded by contract NAS5-30720. We thank M. Nowak, who provided the grouping values used in the XIS spectra. S. Suchy thanks GSFC and UMBC for their hospitality during his visit. 
\end{acknowledgements}

\bibliographystyle{apj}
\bibliography{mnemonic,bibtex_library}

\end{document}